\documentclass[superscriptaddress,preprintnumbers,amssymb,nofootinbib]{revtex4-2}

\usepackage{tikz-feynman}
\usepackage{subfig}
\usepackage{bm}
\usepackage{esvect}
\usepackage{verbatim}
\usepackage{multirow}
\usepackage[left]{lineno}
\usepackage{blindtext}
\pdfoutput=1
\usepackage[T1]{fontenc}
\usepackage{graphicx}
\usepackage{epsfig}
\usepackage{amsmath}
\usepackage{amsfonts}
\usepackage{amssymb}
\usepackage{braket}
\usepackage{cancel}
\usepackage{pstricks}
\usepackage{color}
\usepackage{feynmf}
\usepackage[normalem]{ulem}
\usepackage{epstopdf}
\usepackage{soul}
\usepackage{cancel}

\definecolor{ao(english)}{rgb}{0.0, 0.5, 0.0}

\usepackage{braket}
\usepackage{cancel}
\usepackage{setspace}
\usepackage{pstricks}
\usepackage{xcolor}
\usepackage{float}
\usepackage{mathtools}
\usepackage{multirow}
\usepackage{booktabs}
\usepackage{enumitem}

\def\missET {{\not\!\! E_T}}

\def\gsim{\lower0.5ex\hbox{$\:\buildrel >\over\sim\:$}}
\def\lsim{\lower0.5ex\hbox{$\:\buildrel <\over\sim\:$}}

\def\be{\begin{equation}}
\def\ee{\end{equation}}
\def\bea{\begin{eqnarray}}
\def\eea{\end{eqnarray}}

\def\op{Q}

\begin{document}

\title{Flavor physics at the EIC 
with b-jet tagging}

\author{Shaouly Bar-Shalom}
\email{shaouly@gmail.com}
\affiliation{Physics Department, Technion--Institute of Technology, Haifa 3200003, Israel}
\author{Jose Wudka}
\email{jose.wudka@ucr.edu}
\affiliation{Physics Department, University of California, Riverside, CA 92521, USA}

\date{\today}

\begin{abstract}
We employ an approximate conserved quantum number (defined as "$b$-Parity" in \cite{our_bP_paper}) of the Standard Model (SM):   $b_P=(-1)^n$, where $n$ is the number of produced $b$-jets in the reaction $e + p/A \to n \cdot j_b +X$, to explore new TeV-scale flavor-changing interactions involving the 3rd generation quarks at the EIC; simply by counting the number of $b$-jets in the final state. 
In particular, the  SM
single and di-jet production at 
the EIC which occur through charge current interactions, $e + p/A \to j + \missET$ and $e + p/A \to 2\cdot j + \missET$,  
are $b_P$-even since the $b_P$-violating (i.e, $b_P=-1$) SM signals for these processes 
are necessarily CKM suppressed and, therefore, have a vanishingly small production rate. 
In contrast, new flavor 
physics can generate $b_P=-1$ signals at the EIC whose only significant 
SM background is due to $b$-jet misidentification. We thus show that 
$b_P$ 
can be used as 
a simple and sensitive probe of new flavor violating physics; specifically, we 
 find that counting single $b$-jet events in 
 $e + p/A \to j + \missET$ at 
the EIC with a center-of-mass (CM) energy of $\sqrt{s} \sim 140$ GeV, can probe scales of new physics up to $\Lambda \lsim {\cal O}(5)$ TeV for a certain type of new chiral flavor-changing physics in 3rd generation interactions. This is remarkably more than 30 times larger than the assumed EIC CM energy and it critically depends on the $b$-tagging efficiency and purity as well as the feasibility of electron-beam polarization. 
%
The sensitivity of the di-jet process, $e + p/A \to 2j + \missET$, to these type of new physics is reduced compared to the single-jet channel.
\end{abstract}

\maketitle
\flushbottom

\newpage 

\section{Introduction \label{sec:intro}}

Despite its enormous success, the Standard Model (SM) is widely regarded as a low-energy effective theory of a more fundamental framework. The nature of this underlying physics is currently being explored at the LHC and will be further probed by next-generation colliders.
The Electron-Ion Collider (EIC) at the Brookhaven National Laboratory \cite{EIC1,EIC2}, which is expected to start taking data during the next
decade, is probably going to be the only new high-energy collider worldwide in the next twenty–thirty years, 
so that exploiting its potential to address fundamental questions in particle physics is warranted.  
Indeed, in spite of the fact that the EIC is primarily designed for exploration of nuclear physics \cite{EIC1,EIC2}, there has been a growing interest in pointing out the EIC's sensitivity and capability for precision tests of the SM \cite{1302.6263,Kumar:2016mfi,1612.06927,2107.02134,2112.07747,2204.07557} and for the search of 
Beyond the SM (BSM) physics \cite{1006.5063,1401.6199,2004.00748,2102.06176,2112.02477,2112.04513,2203.01510,2203.13199_Snowmass,2204.07557,2207.10261,2210.09287,2301.02304,2307.00102,2310.08827,2401.08419,2402.17821,2411.13497,2503.02605,2505.08871,2507.02039,2507.21477,2512.15865,2601.00068}. We note, in particular, that the potential high polarization (expected to approach 80\%) of the incoming electron beam at the EIC and its anticipated high integrated luminosity of ${\cal L} \sim {\cal O}(100)$ fb$^{-1}$ is highlighted in many of these studies; for comparison we recall that the HERA $ep$ collider collected ${\cal O}(1)$ fb$^{-1}$ of data and ran with about $30\%$ of average electron beam polarization during its phase II \cite{HERA1,HERA2}.  

In this paper we would like to promote the possibility of exploiting the EIC as a testing platform for the study of flavor-changing (FC) physics, which remains one of the fundamental unresolved matters in theoretical particle physics; a signal of TeV-scale FC interactions will be a clear and unambiguous evidence for BSM physics. There are reasons to believe that FC interactions involving the 3rd generation quarks are likely to show the largest effect.\footnote{Due to the very large number of investigations in this direction that have been carried during the past several decades, an exhaustive citation list cannot be practically provided.}
Following the idea introduced by the authors in \cite{our_bP_paper}, we propose here to test FC new physics (NP) effects in single and di-jet production at the EIC  by simply counting the number of $b$-jets in the final state (FS). As outlined in \cite{our_bP_paper}, this approach is best suited for lepton colliders but in some cases it may be extended to hadron colliders or electron-proton colliders if the $b$-quark content in the proton can be ignored. In particular, we will show in this work that it can be applied to the EIC in charge-current (CC) interactions, by achieving a high purity of the $b$-jet sample and leveraging the polarization of the incoming electron beam. 

Indeed, the combination of high luminosity, high $b$-jet identification efficiency, the availability of polarized beams and the relatively low CM energy of the EIC (below the top-quark threshold) prove combined virtues leading to a rather impressive sensitivity of this collider to the kind of flavor physics investigated below. In particular,  the expected EIC sensitivity  is significantly better to that of HERA  despite the fact that HERA operated at more than twice the CM energy. 

The physics behind the method we use is based on the observation that in the limit where the CKM parameters  $V_{3j}=V_{j3}=0$ ($j\neq 3$), the SM acquires an additional global $U(1)_b$ symmetry ("bottomness"), where $b$ and $\bar b$ have opposite charges, and which holds to any order in perturbation theory. Specifically, since $V_{j3},V_{3j}$ ($j \neq 3$), though not zero, are small [$ V_{31},V_{13} \sim {\cal O}(\lambda^3)$ and $ V_{32},V_{23} \sim {\cal O}(\lambda^2)$, where $\lambda \sim 0.22$ is the Wolfenstein parameter], the quantum number associated with $U(1)_b$ is approximately conserved (in particular, the top quark decays almost exclusively into $ bW$). Then, for example, given a reaction of the type
\begin{eqnarray}
    n_i \cdot b + X \to n_f \cdot b + Y ~, \label{bp1}
\end{eqnarray}
where {\it (i)} $X,Y$ denote sets of particles not containing  $b$-quark/jets;  {\it (ii)} $n_i,n_f$ are the number of $b$-quark/jets in the initial and final states, respectively; and 
{\it (iii)} there are no top quarks in the initial sate, and those in the final state have decayed via $t \to bW$;  we have  $(-1)^{n_i} = 
(-1)^{n_f}$ and we can define an approximately conserved flavor number for collider scattering processes, $b_P \equiv (-1)^{n_f -n_i}$, which was named $b$-Parity in \cite{our_bP_paper}.
Therefore, if there is no $b$-quark content in the initial state ($n_i=0$), then an odd number of $b$-quark jets {\it cannot} be generated in the SM in scattering processes, as was pointed out in \cite{our_bP_paper}.

In this work we wish to apply the concept of $b$-Parity for searches  of new BSM sources of FC physics in the third generation quark sector at the EIC, in multi-jet production:
\begin{eqnarray}
e + p/A \to n \cdot j_b + m \cdot j_\ell + X \label{nbx}~,
\end{eqnarray}
where $n$ denotes the number of $b$ and/or $\bar b$ jets in the FS,\footnote{The method used here does not require differentiating between $b$ and $\bar b$-quark jets.} $m$ is the number of light-jets and $X$ stands for 
leptons and/or missing energy. 
We assume below an EIC running phase with a CM energy $\sqrt{s} < m_t$, so that the production of
$b$-quarks via $t \to b W$ is vanishingly small, as it will occur only through an off-shell top-quark in the FS. 
Thus, to the extent that the $b$-quark content inside the proton can be ignored, the $b$-Parity number for the reaction in Eq.\ref{nbx} becomes: 
\begin{eqnarray}
b_P=(-1)^n ~, \label{bParity}
\end{eqnarray}
so that the measured quantum number reduces to the net number of detected $b$-quark jets, in which case 
it is convenient to use the derived quantity $ b_P $. 
The only SM processes that violate $b_P$ in these multi-jet production processes at the EIC necessarily involve the CC $u \to b$ and/or $c \to b$ interactions, which are suppressed by the corresponding small off-diagonal CKM element factors, $|V_{ub}|^2$ and $|V_{cb}|^2$, so that the SM is $b_P$-even at the EIC to a good accuracy.
Thus, to experimentally detect 
 NP signals with $b_P=-1$ in the processes of the type shown in Eq.~\ref{nbx}, one should simply measure/count the number of events with an odd number of $b$-jets in the FS.

Denoting by $t_j$ the light-jet 
mis-tagging probability (i.e., that of mistaking a light-quark jet or gluon-jet for 
a $b$-jet)\footnote{We do not differentiate between a light-jet and a $c$-quark jet, since the contribution of $c$-quark jets in the reactions discussed below (although included) is negligible. In particular, the single and di-jet SM processes which lead to a $c$-jet in the final state contribute less than $1\%$ in both these channels and, therefore, have a negligible effect on our analysis in the next sections.} and by $\epsilon_b$ the $b$-tagging efficiency, 
the probability (or cross-section) for detecting precisely $k$ $b$-jets 
in the reaction of Eq.\ref{nbx} is given by
\begin{eqnarray}
\bar\sigma_{kj_b} = \sum_{u,v} P_u^n P_v^m 
\left[ \epsilon_b^u (1-\epsilon_b)^{n-u} \right] 
\left[ t_j^v (1-t_j)^{m-v} \right] \sigma_{ n m} \delta_{u+v,k} 
\label{sigmabar} ~,
\end{eqnarray}
where $P^i_j = i!/j!/(i-j)!$ and $\sigma_{ n m} = \sigma (e +p/A \to n \cdot j_b + m \cdot j + X)$.

Clearly, the main obstacle in the search for $b_P$-odd NP signals is, therefore, the reducible SM
background due to jet mis-identification, which results
from having a non-optimal $b$-tagging efficiency ($\epsilon_b$) below 1, 
and/or having non-zero probabilities ($t_j$) of mis-tagging a light-quark jet for a $b$-jet (also known as the purity of the $b$-jet sample). This type of background would of course disappear as $\epsilon_b \to 1$ and $t_j \to 0$, but even 
with the seemingly small $t_j \sim 0.01$ (and a relatively high $b$-tagging efficiency of $\epsilon_b \sim 0.8$) can produce a significant number of 
(miss-identified) "fake" $b_P$-odd events in the detector. 

In what follows, we will consider new flavor physics involving the third generation quarks, which impacts single $b$-jet production at the EIC. We use effective field theory to describe the NP and find that the EIC will be sensitive to NP scales of $\Lambda \lsim 1-3 $ TeV and for a certain type of interaction up to 
$\Lambda \lsim 5$ TeV. As will be shown, this will require a relatively high purity of the $b$-jets sample and a substantial polarization of the electron-beam. 

\section{new 3rd generation flavor physics and the EIC \label{sec:3rdflavor}}

With no indications of NP observed so far up to the TeV scale, it 
is useful to parameterize
the impact of new heavy states by higher dimensional, gauge-invariant effective operators, $\op_i^{(n)}$, in the so-called SM Effective Field Theory (SMEFT) framework, where the effective operators are constructed using the SM fields and their coefficients are suppressed by inverse powers of the NP scale $\Lambda$
\cite{EFT1,EFT2,EFT3,EFT4,EFT5}:
\begin{eqnarray}
{\cal L} = {\cal L}_{SM} + \sum_{n=5}^\infty
\frac{1}{\Lambda^{n-4}} \sum_i \alpha_i \op_i^{(n)} \label{eq:EFT1}~,
\end{eqnarray}
where here $n$ is the mass dimension of $\op_i^{(n)}$ and we assume decoupling and weakly-coupled heavy NP, so that
$n$ equals the canonical dimension. 
One can further divide the higher-dimension effective operators into those that can be potentially generated at tree-level (PTG) 
and those that are necessarily loop generated (LG) by the underlying heavy theory \cite{jose_PTG_LG}.   
The dominating NP effects are then expected to be generated by contributing operators with the 
lowest dimension (smallest $n$) that are PTG in the underlying UV theory. 
The (Wilson) coefficients $\alpha_i$ depend on the details and dynamics of the underlying heavy theory and, therefore, 
they parameterize all possible weakly-interacting and decoupling types of heavy physics. In particular, 
it is expected that  $\alpha_i =O(1)$ for the PTG operators (for favorable types of NP), and  $\alpha_i \sim 1/(4\pi)^2$ for all LG operators; the effects of LG operators are thus a-priory suppressed by a loop factor and, therefore, their effects at lower energies (i.e., $E < \Lambda$) are expected to be subleading. 

We will use below a more practical "effective" NP scale, which is the variable that enters the calculation: 
\begin{eqnarray}
    \Lambda_{\tt eff} \equiv \frac{\Lambda}{\sqrt{\alpha}} ~; \label{lam_eff}
\end{eqnarray}
%
note that $\Lambda_{\tt eff} \sim \Lambda$ for natural PTG couplings. 

As mentioned above, in this work we characterize and exploit the possible manifestations of new flavor structures which involve the 3rd generation quarks, assuming a generic decoupling and weakly-coupled underlying heavy NP and using the SMEFT framework. In particular, we examine below 
single and di-jet production at the EIC via $2 \to 2$ and $2 \to 3$ processes, focusing (for reasons explained below) on CC interactions comprising a single $b$-quark in the FS. The relevant dimension six (dim.6) operators that are PTG and that can mediate sizable new flavor physics effects in these reactions involve a FC $b \to u$ transition and are listed in Table \ref{tab:1}.

Let us make a few comments on the effective operators listed in Table \ref{tab:1}:
\begin{itemize}
    \item We focus only on the operators that mediate $b \to u$ flavor transitions, generating the 
    $e \nu_ebu$ 4-Fermi contact terms and the $Wub$ charged currents, since these yield the dominant contribution to single $b$-jet production at the EIC, e.g., via $e u \to \nu_e b$ and $e u \to \nu_e b g$ (see next section). In particular, we find that the sensitivity to the corresponding operators which generate $b \to c$ transitions is significantly reduced, due to the 
    suppressed $c$-quark content inside the proton.  
    \item The operators $\op_{\ell e dq}$, $\op_{\ell e qu}^{(1)}$ and
    $\op_{\ell e qu}^{(3)}$ are non-hermitian and no symmetry applies to their indices, so that, for example, the $(1131)$ entry is different from the $(1113)$ one.\footnote{The numbers denote the generation of the fermions as they appear in the operator label.} We note that, for the CC reactions $e + p/A \to j_b + \nu_e$ and $e + p/A \to j_b + j_\ell + \nu_e$, only the $(1131)$ entries of these operators contribute.
    \item The operator $\op_{Hud}$ is also non-hermitian and $\alpha_{Hud}(31) \neq \alpha_{Hud}(13)$; here also only $\op_{Hud}(13)$ contributes to $e + p/A \to j_b + \nu_e$ and $e + p/A \to j_b + j_\ell + \nu_e$. 
    \item The operators $\op_{\ell q}^{(3)}$ and $\op_{Hq}^{(3)}$ are hermitian so that, for real Wilson coefficients: 
    $\alpha_{\ell q}^{(3)}(1131) = \alpha_{\ell q}^{(3)}(1113)$ and $\alpha_{Hq}^{(3)}(13) = \alpha_{Hq}^{(3)}(31)$.
\end{itemize}

%
\begin{table*}[htb]
\caption{
The dim.6 effective operators that are PTG in the underlying heavy theory and that can mediate the CC interaction 
$e + p/A \to j_b + \nu_e$ and $e + p/A \to j_b + j_\ell + \nu_e$, where 
$\ell_1,q_1$ are the lepton and quark SU(2) left-handed doublets of the first generation and $e,u,d,b$ are the electron and $u,d,b$-quarks right-handed SU(2) singlets; $j,k$ are SU(2) indices. 
We list the chirality structure of the fermionic currents, possible types of underlying heavy physics that can generate these operators and the additional $\ell \ell q q^\prime$ contact terms that are generated by the 4-Fermi operators due to SU(2) gauge invariance. \label{tab:1}}
\renewcommand{\arraystretch}{1.5}
\begin{tabular}[t]{c|c|c}
effective operator & Chirality (type) & Interactions  \\
\hline
$\op_{\ell q}^{(3)}(1113) = \left(\bar \ell_1 \gamma_\mu \tau^I \ell_1 \right) \left(\bar q_1 \gamma^\mu \tau^I q_{3} \right)$  & 
$(\bar L L) (\bar L L)$ (vector) & $eetu$, $eebd$, $e\nu_e b u$ \\
$\op_{\ell e q u}^{(1)}(1131) = \left(\bar \ell_1^j e \right) \epsilon_{jk} \left(\bar q_3^k  u \right)$  & 
$(\bar L R) (\bar L R)$ (scalar,tensor)  & $eetu$, $e \nu_e b u$  \\
$\op_{\ell e q u}^{(3)}(1131) = \left(\bar \ell_1^j \sigma_{\mu \nu} e \right) \epsilon_{jk} \left(\bar q_3^k  \sigma_{\mu \nu} u \right)$ & 
$(\bar L R) (\bar L R)$ (scalar,tensor) & $eetu$, $e \nu_e b u$  \\
$\op_{\ell e dq}(1131) = \left(\bar \ell_1^j e \right) \left(\bar b  q_1^j \right)$   & 
$(\bar L R) (\bar R L)$ (scalar,vector) & $eebd$, $e \nu_e b u$  \\
$\op_{Hud}(13) = i \left(\tilde H^\dagger D_\mu H \right) \left(\bar u  \gamma^\mu b \right)$ +h.c.   & 
$(\bar R R) $ (vector,fermion) & $W u b$  \\
$\op_{Hq}^{(3)} (13) = i \left(H^\dagger i \stackrel{\leftrightarrow}{D^I}_\mu H \right) \left(\bar q_1  \gamma^\mu \tau^I q_3 \right)$   & 
$(\bar LL ) $ (vector,fermion) & $W u b$ 
\end{tabular}
\end{table*}

Bounds on the operators listed in Table \ref{tab:1} can be derived from Drell-Yan (DY) processes at the LHC, via the CC $\bar b u \to e^+ \nu_e$ and neutral current (NC) $\bar b  d \to e^+  e^-$ interactions \cite{Greljo:2022jac,Allwicher:2022gkm,Greljo:2023bab,Grunwald:2023nli,Hiller:2025hpf}, from the $b$-decays $b \to u e \nu_e$ and $b \to d ee$ \cite{Greljo:2022jac,Greljo:2023bab} and from $e^+ e^- \to t \bar u$ at LEP2 \cite{Bar-Shalom:1999dtk,L3,DELPHI} and $t \bar t$ production at the LHC followed by the FC top decay $t \to u e^+ e^-$ 
\cite{Maltoni-global,topdecay1,1008.3562,topdecay3,topdecay2}.   
While the latter (i.e., top-quark production and decays) gives the weakest bounds 
of $\Lambda_{\tt eff} \gsim {\cal O}(1)$ TeV, analysis of the DY processes at the LHC mentioned above yield $\Lambda_{\tt eff} \gsim 3-7$ TeV, depending on the type of operator and process. The bounds from $b$-decays are the strongest, reaching  in some cases 
$\Lambda_{\tt eff} > {\cal O}(10-20)$ TeV when a single NP operator is turned on. As mentioned earlier and will be shown below, the sensitivity to these dim. 6 effective operators at the EIC strongly depends on the $b$-tagging efficiency and purity and on the availability of polarized electron beams, where, for most operators, it is expected to be more sensitive to this type of NP than the top-quark systems (mentioned above) and, for a certain type of 4-fermi interaction, it is competitive with the current bounds from DY processes at the LHC. 

\section{$b$-Parity and single jet production at the EIC \label{sec:bP}}

Let us consider $2 \to 2$ processes from CC and NC interactions, which are potentially the most sensitive to new flavor physics:
\begin{eqnarray}
    {\rm CC:} ~~~ && e + p/A \to j + \missET ~, \label{CCeq} \\
    {\rm NC:} ~~~ && e + p/A \to j + e ~, \label{NCeq}
\end{eqnarray}
where $j=j_b$ and $j=j_l$ for a $b$-jet and light-quark jet, respectively, and 
$\missET$ stands for missing energy from the outgoing electron-neutrino $\nu_e$. 

In the limit of a diagonal CKM, the CC parton-level process in Eq.\ref{CCeq} proceeds in the SM only via the t-channel $W$-boson exchanges  
$e u \to d \nu_e$ (this channel contributes more than $90\%$ to the total CC $2 \to 2$ cross-section in the SM), $e c \to s \nu_e$,  
$e \bar d \to \bar u \nu_e$ and $ e \bar s \to \bar c \nu_e $, while the NC process in Eq.\ref{NCeq} proceeds via the t-channel $\gamma$ and $Z$-boson exchanges 
$e q \to q e $ and 
$e  \bar q \to \bar q  e $, where $q=u,d,c,s,b$ (in this case $e  u \to e u $ and $e  d \to e d $ contribute about $80\%$ and $10\%$, respectively, of the total NC $2 \to 2$ SM cross-section). 

Note that a single $b$-jet can be produced in the SM via the NC reaction in Eq.\ref{NCeq}, even in the limit of a diagonal CKM mixing matrix, due to the presence of the $b$-quark in the proton via $e b \to e b $ and $e  \bar b \to e \bar b $; this however accounts for a marginal fraction (about 0.3\%) of the total $2 \to 2$ NC cross-section. The total SM contribution to the 
single $b$-jet signal via the NC process of Eq.\ref{NCeq} is driven primarily by light jets mis-tagged as $b$-jets, and we find that it is overwhelmingly large relative to the NP effects under consideration (see next section), even for a  high purity $t_j \sim 0.001$ of the $b$-jet sample.
In other words, the SM portion of the single $b$-jet signal in the NC process case of 
Eq.\ref{NCeq} (due to $t_j \neq 0$) dominates our $b_P$ analysis, to the level that it cannot yield useful bounds on the new flavor physics involving the $b$-quark.\footnote{In the $2 \to 2$ NC case of single jet production ($e + p/A \to j + e$, see Eq.~\ref{NCeq}) we find that the SM effective cross-section (i.e., the SM background) is about three orders of magnitude larger than the NP one even with $t_j \sim 0.001$: $\bar\sigma_{1j_b}^{NC}(SM) \sim {\cal O}(1000~{\rm fb})$ while $\bar\sigma_{1j_b}^{NC}(NP) \sim {\cal O}(1~{\rm fb})$, for $t_j \sim 0.001$ and $\epsilon_b \sim 0.8$. Thus, the expected 95\% CL bound on $\Lambda_{\tt eff}$ in this case (see Eq.~\ref{nsd}) is of 
${\cal O}(100~{\rm GeV})$ and, therefore, not useful.}

We therefore focus in the rest of this study on the CC process of Eq.\ref{CCeq} and define the  
corresponding effective SM and NP cross-sections for producing a single $b$-jet, that we denote by
$\bar\sigma_{1j_b}$ (cf. Eq.\ref{sigmabar}) , which takes into account the non-ideal $b$-jet tagging efficiency and purity:\footnote{We neglect the effects of SM$\times$NP interference for the processes being considered since it is CKM suppressed.}
\begin{eqnarray}
    \bar\sigma_{1j_b}^{CC}(SM) &=& t_j \cdot \sigma_{SM}^{CC} ~~;~~ \sigma_{SM}^{CC} = \sigma(e + p/A \to j_l + \missET) ~, \label{sigeffSM}\\ 
    \bar\sigma_{1j_b}^{CC}(NP) &=& \epsilon_b \cdot \sigma_{NP}^{CC} ~~; ~~ \sigma_{NP}^{CC} = \sigma(e + p/A \to j_b + \missET) ~,\label{sigeffNP}
\end{eqnarray}
so that, as noted above, the SM contribution to $e + p/A \to j_b + \missET$ arises due to mis-tagged light-quark jets as $b$-jets in $e + p/A \to j_l + \missET$. We find that $\sigma_{SM}^{CC} \sim 13$pb at the EIC with beam energies of $E_e=18$ GeV and $E_p=275$ GeV (see also below), while the irreducible SM background from $e + p/A \to j_b + \missET$ arises from the parton level $eu \to b \nu_e$ and $ec \to b \nu_e$ processes and, therefore, is CKM suppressed and negligibly small compared to $\sigma_{SM}^{CC}$. In particular, $\sigma_{SM}(e u \to b \nu_e) \sim \sigma_{SM}(e u \to d \nu_e) \times |V_{ub}|^2$ and $\sigma_{SM}(e c \to b \nu_e) \sim \sigma_{SM}(e c \to s \nu_e) \times |V_{cb}|^2$ and these two channels yield 
an overall single $b$-jet + MET cross-section of $\sigma_{SM}(e + p/A \to j_b + \missET) \sim 5 \cdot 10^{-4}$ pb, which is more than four orders of magnitudes smaller than $\sigma_{SM}^{CC}$ in Eq.~\ref{sigeffSM}.

Another potential SM (reducible) background may arise from the $2 \to 2$ NC process $e + p/A \to j + e$, due to non-ideal detector performance and event selection, when the electron is missed; in this case such an event appears as a CC event, i.e., with a single jet and MET in the final state. We note, however, that this type of background is expected to be easier to control at the EIC (compared e.g., to the LHC), since its initial state energy is relatively well known and so global energy-momentum conservation is easier to apply. Also, the electron-Proton/Ion Collider (ePIC) detector is specifically designed with a very good forward coverage specifically for the scattered electron as well as a very good MET resolution (i.e., about an order of magnitude better than at the LHC), so that lost-electron backgrounds producing fake MET are expected to be much less problematic at the EIC, see \cite{EIC1} and subsequent ePIC studies; a detailed study of such detector simulations of fake MET signals is beyond the scope of this paper.

We, therefore, do not include below the SM background from the CKM suppressed channels and from fake MET signals due to missed electrons in the NC channel $e + p/A \to j + e$.
In particular, for the total CC $j_b + \missET$ production cross-section we take:
\begin{eqnarray}
    \bar\sigma_{1j_b}^{CC}(\epsilon_b,t_j,\Lambda_{\tt eff})=\bar\sigma_{1j_b}^{CC}(SM) + \bar\sigma_{1j_b}^{CC}(NP) 
  ~,\label{sigeff}
\end{eqnarray}
which depends on the $b$-tagging and purity factors as well as on the effective NP scale, $\Lambda_{\tt eff}$ (see Eq.\ref{lam_eff}). 
The projected performance for the tagging efficiencies at the EIC is still not known and, in the following, we will use values in the ranges $0.6<\epsilon_b<0.8$ and, depending on $\epsilon_b$, $0.001 < t_j < 0.1$. Specifically, we consider below three representative benchmark choices for the pair $(\epsilon_b,t_j)$, spanning  from optimistic to conservative scenarios. We will also present in some instances results as a function of $t_j$
over a broader interval, since, as discussed below, this parameter plays a central role in the analysis performed here. We hope that these setups will provide a challenging yet interesting cases to investigate at the EIC.

In Table \ref{tab:CSX} we list the NP cross-sections $\sigma_{NP}^{CC} = \sigma(e + p/A \to j_b + \nu_e)$, corresponding to each of the operators in Table \ref{tab:1}, with beam energies $E_p=275$ GeV and $E_e= 18$ GeV, i.e., a CM energy of $\sqrt{s} = 141$ GeV. 
All cross-sections (here and in the next sections) were calculated using {\sc MadGraph5\_aMC@NLO}~\cite{madgraph5} at LO parton-level and with the SMEFTsim model of~\cite{SMEFTsim1,SMEFTsim2} for the EFT framework. The 5-flavor scheme was used, with the CT18NNLO parton distribution functions \cite{1912.10053} and the default {\sc MadGraph5\_aMC@NLO} LO dynamical scale. Also, the acceptance cuts of $p_{T_j} > 20$ GeV and $|\eta_j| < 3.5$ were applied on the transverse momentum and pseudo-rapidity distributions of the light-quark and/or $b$-quark jets in the final state.
\begin{table*}[htb]
\caption{The NP cross-sections [fb] at the EIC for the CC production of a single $b$-quark, $\sigma_{NP}^{CC} = \sigma(e + p/A \to j_b + \nu_e)$, that each of the  operators in Table \ref{tab:1} yield. The cross-sections are calculated with beam energies $E_p=275$ GeV and $E_e= 18$ GeV, see also text. \label{tab:CSX}}
\renewcommand{\arraystretch}{1.5}
\begin{tabular}[t]{c|c|c|c|c|c|c}
& $\op_{\ell q}^{(3)}(1113)$ & $\op_{\ell e q u}^{(1)}(1131)$ &
 $\op_{\ell e q u}^{(3)}(1131)$ & $\op_{\ell e dq}^{(1)}(1131) $ & 
 $\op_{Hud}(13)$ & $\op_{Hq}^{(3)}(13)$ \\
\hline
$\sigma(e + p/A \to j_b + \nu_e)$ & 76.4  & 1.5  & 176 & 1.5  & 4.2  & 45.4  
\end{tabular}
\end{table*}

To asses the sensitivity of the $b_P$-odd 
NP signal of our interest, we use the number of expected single $b$-jet events, which correspond to $\bar\sigma_{1j_b}^{CC}(\epsilon_b,t_j,\Lambda_{\tt eff})$ in Eq.\ref{sigeff}:
\begin{eqnarray}
    N_{1j_b}^{CC}(\epsilon_b,t_j,\Lambda_{\tt eff}) = {\cal L} \cdot {\cal A} \cdot \bar\sigma_{1j_b}^{CC}(\epsilon_b,t_j,\Lambda_{\tt eff}) ~, \label{NCC} 
\end{eqnarray}
and we will henceforward take ${\cal L}=100$ fb$^{-1}$ as the projected integrated EIC luminosity and ${\cal A}$ as the overall acceptance + efficiency factor which we will set below to ${\cal A}=0.8$.   

The sensitivity is then determined  
by comparing the theoretical shift due to the underlying 
$b_P$-odd interactions with the expected error ($\Delta$) 
in measuring the given quantity. 
Thus,
requiring a signal of at least $N_{SD}$ standard deviations, we have
\begin{eqnarray}
\left| N_{1j_b}^{CC}(\epsilon_b,t_j,\Lambda_{\tt eff}) - N_{1j_b}^{CC}(SM) \right| \geq N_{SD} \cdot \Delta \label{nsd}~.
\end{eqnarray}
where $N_{1j_b}^{CC}(SM) = {\cal L} \cdot {\cal A} \cdot \bar\sigma_{1j_b}^{CC}(SM)$ (see Eq.\ref{sigeffSM}) is the expected number of $e + p/A \to j_b + \missET$ events in the SM due to light-jet mis-identification as $b$-jets.
We include three contributions in the overall expected error $ \Delta $: 
\begin{description}
    \item[Statistical error] $\Delta_{\rm stat}=\sqrt{N_{1j_b}^{CC}(\epsilon_b,t_j,\Lambda_{\tt eff})}$ 
\item[Systematic error] $\Delta_{\rm sys}=N_{1j_b}^{CC}(\epsilon_b,t_j,\Lambda_{\tt eff}) \cdot \delta_s$ 
\item[Theory error] $\Delta_{\rm theor}= N_{1j_b}^{CC}(\epsilon_b,t_j,\Lambda_{\tt eff})
\cdot \delta_t$
\end{description} 
which we combine in
quadrature: $\Delta^2 = \Delta_{\rm stat}^2 + \Delta_{\rm sys}^2 + \Delta_{\rm theor}^2$, where the  $ \delta_{s},\delta_{t} $ denote the statistical and theoretical errors per event; $ \delta_s $ is usually estimated using experimental values from related processes and $ \delta_t $ is derived from the errors in the Monte Carlo integration used in calculating the various cross sections. Unless stated otherwise, we will use below 
$ \delta_{s} = \delta_{t} = 0.02 $.  
\begin{figure}[htb]
  \centering
\includegraphics[width=0.4\textwidth]{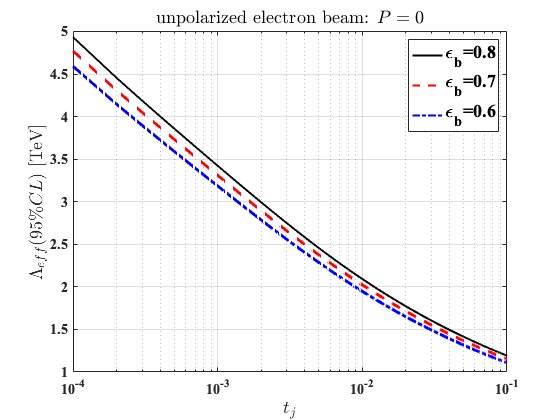}
\includegraphics[width=0.4\textwidth]{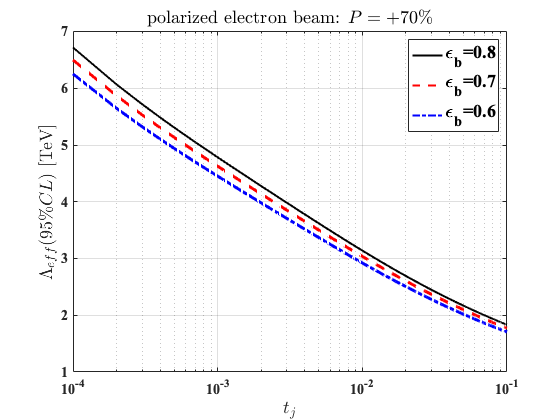}
\caption{Expected 95\% CL bounds on $\Lambda_{\tt eff}$ [TeV] from 
the CC single-jet signal, $e + p/A \to j + \missET$, for the Tensor operator $\op_{\ell e q u}^{(3)}(1131) $, as a function of the 
light-jet mis-tagging purity factor $t_j$, 
for three values of the $b$-tagging efficiency $\epsilon_b=0.6,0.7,0.8$. On left: without electron beam polarization and on right: with $P_e=0.7$, i.e, $+70\%$ right-handed electron beam polarization.}
  \label{fig:1}
\end{figure}
\begin{table*}[htb]
\caption{\label{tab:2}
The expected 95\% CL bounds on the effective scale $\Lambda_{\tt eff} = \Lambda/\sqrt{\alpha}$ for the operators in Table \ref{tab:1}. Numbers are given for the three $b$-tagging efficiency setups: "loose", "medium" and "tight" and for unpolarized and $+70\%$ ($P_e=+0.7$, see Eq.~\ref{sig_pol}) polarized incoming electron beam. See also text. \label{tab:2}}
\renewcommand{\arraystretch}{1.5}
\begin{tabular}[t]{c|c|c|c}
& \multicolumn{3}{c}{95\% $CL$ bounds on $\Lambda_{\tt eff}(\op)$ [TeV]} \\
& \multicolumn{3}{c}{with electron beam polarization $P_e=0~(+0.7)$} \\
\hline 
& "tight" setup & "medium" setup & "loose" setup \\
$\op$ & $(\epsilon_b,t_j)=(0.6,0.001)$  & $(\epsilon_b,t_j)=(0.7,0.01)$ & $(\epsilon_b,t_j)=(0.8,0.03)$ \\
\hline
$\op_{\ell q}^{(3)}(1113)$  & 2.6 (2.4) & 1.7 (1.6) & 1.3 (1.3) \\
$\op_{\ell e q u}^{(1)}(1131)$  & 1.0 (1.4)  & 0.6 (0.9) & 0.5 (0.8) \\
$\op_{\ell e q u}^{(3)}(1131) $ & 3.2 (4.6) & 2.0 (3.1) & 1.6 (2.5) \\
$\op_{\ell e dq}^{(1)}(1131) $   & 1.0 (1.4) & 0.6 (0.9) & 0.5 (0.8)\\
$\op_{Hud}(13) $  & 1.3 (1.2) & 0.8 (0.7) & 0.6 (0.6) \\
$\op_{Hq}^{(3)} (13) $   & 2.3 (2.1) & 1.5 (1.4) & 1.2 (1.1) 
\end{tabular}
\end{table*}

In Fig.~\ref{fig:1} we plot the 95\% CL bounds on $\Lambda_{\tt eff}$ for the tensor operator $\op_{\ell e qu}^{(3)}(1131)$, 
as a function of the light-jet mis-tagging purity factor $t_j$ and for three values of the
$b$-jet tagging efficiency
$\epsilon_b=0.6,0.7,0.8$. 
In Table \ref{tab:2} we give the 95\% CL bounds on $\Lambda_{\tt eff}$ for all the operators in Table \ref{tab:1}, where 
we consider three representative tagging efficiency/purity scenarios:
"loose": $(\epsilon_b,t_j)=(0.8,0.03)$, 
"medium": $(\epsilon_b,t_j)=(0.7,0.01)$ and
"tight": $(\epsilon_b,t_j)=(0.6,0.001)$. 
We see from both Fig.~\ref{fig:1} and Table \ref{tab:2} that the critical parameter in our case is the purity factor, i.e., the "tight" $b$-tagging setup is by far the most  advantageous. This is indeed expected when the background from light-quark jets is large (as in our case), so that the analysis benefits more from purity rather than from a higher $b$-tagging efficiency.

\subsection{Effects of beam polarization \label{sec:beam_pol}}

Let us next discuss the effects of beam polarization at the EIC. As mentioned earlier, an important feature of the EIC is the possibility of electron beam polarization, which is particularly useful for improving the sensitivity to NP with chiral couplings, as in our case. 
The polarized cross-section is:
\begin{eqnarray}
    \sigma(P_e)=\frac{1}{2} \left[ \left(1-P_e\right) \sigma_{-} + \left(1+P_e\right) \sigma_{+} \right] ~, \label{sig_pol}
\end{eqnarray}
where $P_e$ is the percentage longitudinal polarization, so that $\sigma_{-}=\sigma(P_e=-1)$ and $\sigma_{+}=\sigma(P_e=+1)$ are the purely left-handed and right-handed cross-sections, respectively.  
In Fig.~\ref{fig:1} and Table \ref{tab:2} we also show the expected 95\% CL bounds on $\Lambda_{\tt eff}$ with $P_e=+70\%$, i.e., a 70\% right-handed incoming electron beam. We see that beam polarization will significantly improve the sensitivity to NP contributing to right-handed incoming electrons by up to $\sim 50\%$, since the SM contribution to these single jet events arises from left-handed incoming electrons. 

\begin{figure}[htb]
  \centering
\includegraphics[width=0.4\textwidth]{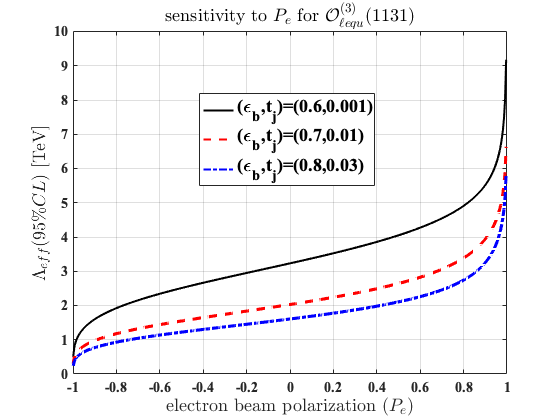}
\includegraphics[width=0.4\textwidth]{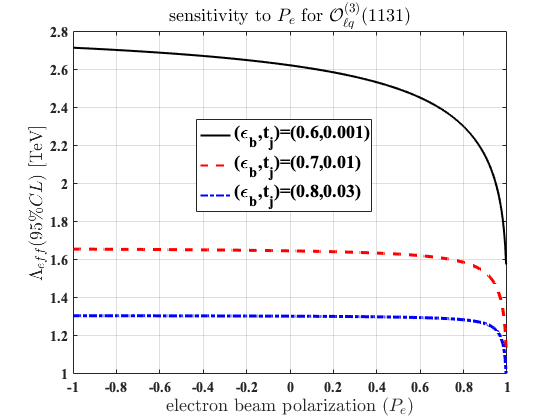}
\caption{Expected 95\% CL bounds on $\Lambda_{\tt eff}$ [TeV] from 
the CC single-jet signal, $e + p/A \to  j + \missET$, as a function of the 
electron beam polarization, $P_e$, for 
the "loose", "medium" and "tight" $b$-tagging scenarios. Results are shown for the tensor operator $\op_{\ell e q u}^{(3)}(1131) $ (left) and the vector operator $\op_{\ell q}^{(3)}(1113) $ (right).}
  \label{fig:3}
\end{figure}

To get a better assessment of the potential effects of electron beam polarization ($P_e$) on the sensitivity to NP with specific chiral couplings,  we depict in Fig.\ref{fig:3} the 95\% CL bound 
on $\Lambda_{\tt eff}$ (denoted by $\Lambda_{\tt eff}(95\% CL)$ in the plot), as a function of $P_e$ for the tensor and vector 4-Fermi operators, $\op_{\ell e q u}^{(3)}(1131) $ and
$\op_{\ell q }^{(3)}(1113) $, in the three "loose", "medium" and "tight" $b$-tagging scenarios. In order to understand the behavior of the curves in Fig.\ref{fig:3}, we recall that for the CC process $ e + p/A \to j_b + \missET$ of our interest here,   
the tensor operator $\op_{\ell e q u}^{(3)}(1131) $ contributes 
to $ e_R + p/A \to j_b + \missET$ (i.e., incoming right-handed electron beam), while the contribution from the vector operator $\op_{\ell q }^{(3)}(1113) $ (as well as the SM background) involve 
left-handed incoming electrons, i.e., $ e_L + p/A \to j_b + \missET$.
In particular, notice that with $P_e \to +100\%$ one would be able to probe $\Lambda_{\tt eff} \to {\cal O}(10)$ TeV for the tensor operator; although not realistic, it is still quite remarkable bearing that the CM energy used for the EIC in this simulation is $E_{CM} \sim 140$ GeV, so that this collider will approach  a sensitivity to $\Lambda_{\tt eff} $ of up to $\sim 70 \times E_{CM}({\rm EIC})$ if $P_e \to +100\%$!

One can further define a polarization asymmetry, which turns out to be a useful sensitive probe of chiral NP effects:
\begin{eqnarray}
{\cal A}_{P_e}^{CC}(\epsilon_b,t_j,\Lambda_{\tt eff}) = \frac{N_{1j_b}^{CC}(\epsilon_b,t_j,\Lambda_{\tt eff};P_e)-N_{1j_b}^{CC}(\epsilon_b,t_j,\Lambda_{\tt eff};-P_e)}
{N_{1j_b}^{CC}(\epsilon_b,t_j,\Lambda_{\tt eff};P_e) + N_{1j_b}^{CC}(\epsilon_b,t_j,\Lambda_{\tt eff};-P_e)} ~, \label{pol_asym}
\end{eqnarray}
where $N_{1j_b}^{CC}(\epsilon_b,t_j,\Lambda_{\tt eff};P_e)$ is the number of expected single $b$-jet $+\missET$ events from initial electron beam with polarization $P_e$ which is extracted from the corresponding polarized cross-section using Eq.\ref{sig_pol}. For the SM (and/or for any type of NP that contributes only to left-handed incoming electron beam, i.e., only to $e_L + p/A \to j + \missET$) we have ${\cal A}_{P_e}^{CC}(SM) = - P_e$, whereas in general we find:
\begin{eqnarray}
    {\cal A}_{P_e}^{CC}(\epsilon_b,t_j,\Lambda_{\tt eff}) = P_e \cdot \frac{\epsilon_b \cdot \sigma_{NP}^{CC} - t_j \cdot \sigma_{SM}^{CC}}{\epsilon_b \cdot \sigma_{NP}^{CC} +  t_j \cdot \sigma_{SM}^{CC}} ~, \label{polA_gen}
\end{eqnarray}
where $\sigma_{SM}^{CC}$ and $\sigma_{NP}^{CC}$ are the cross-sections for the production of $j_l + \missET$ and $j_b + \missET$ at the EIC, respectively, without the tagging efficiencies, see Eqs.\ref{sigeffSM} and Eq.\ref{sigeffNP}; note that $ \sigma_{NP}^{CC} \propto 1/ \Lambda_{\tt eff}^4 \cdot $.

In Fig.\ref{fig:4} we plot the resulting polarization asymmetry for the SM case (i.e., no NP insertion) 
and for the case 
where the tensor operator $\op_{\ell e q u}^{(3)}(1131) $ is turned on with $\Lambda_{\tt eff}=1$ TeV, 
as a function of $P_e$ and for the three "loose", "medium" and "tight" $b$-tagging scenarios. As mentioned above, ${\cal A}_{P_e}^{CC}(SM) = - P_e$, so that the SM case is independent of the $b$-tagging and purity factors. We see that the asymmetry is very sensitive to the purity factor $t_j$ and, in particular, it approaches its SM value as $t_j$ grows. This is due to the growth of the SM background in the single $b$-jet production rate as $t_j $ increases, to the level that it dominates the single $b$-jet + missing energy signal. We find that the sensitivity of ${\cal A}_{P_e}^{CC}(\epsilon_b,t_j,\Lambda_{\tt eff})$ to the operators that mediate 
$e_R + p/A \to j_b + \missET$ (i.e., with a right-handed electron beam) is comparable to what we found above simply by counting single $b$-jet events, cf. Eq.\ref{nsd}.

\begin{figure}[htb]
  \centering
\includegraphics[width=0.6\textwidth]{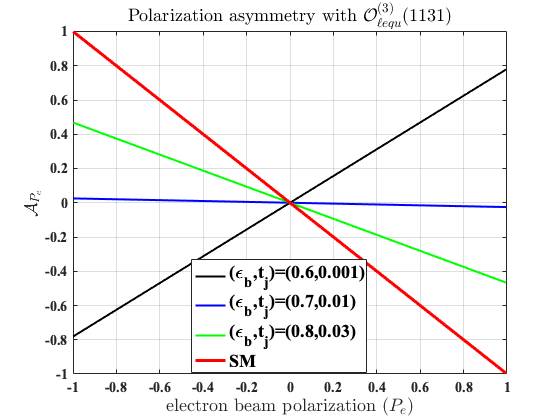}
\caption{The polarization asymmetry 
${\cal A}_{P_e}^{CC}(\epsilon_b,t_j,\Lambda_{\tt eff})$ of Eq.\ref{pol_asym} when the tensor operator $\op_{\ell e q u}^{(3)}(1131)$ is turned on with $\Lambda_{\tt eff}=1$ TeV, as a function of the electron beam polarization $P_e$. The asymmetry is plotted for 
the "loose", "medium" and "tight" $b$-tagging scenarios and also for the SM case. See also text.}
  \label{fig:4}
\end{figure}

\section{$b$-Parity and di-jet production at the EIC \label{sec:bP2}}

The di-jet production at the EIC can also be separated into the CC and NC interactions: 
\begin{eqnarray}
    {\rm CC:} ~~~ && e + p/A \to 2 \cdot j + \missET ~, \label{CCeq2j} \\
    {\rm NC:} ~~~ && e + p/A \to 2 \cdot j + e ~, \label{NCeq2j}
\end{eqnarray}
where 
within the the SM in the limit of a diagonal CKM, the parton-level processes contributing to the CC  reaction in Eq.\ref{CCeq2j} are the electron-gluon fusion $e g \to d \bar u \nu_e ,s \bar c \nu_e $ and
$e  u \to d g \nu_e$, $e c \to s g \nu_e$,  
$e \bar d \to \bar u g \nu_e$, $ e \bar s \to \bar c g \nu_e $; the latter group correspond 
to the $2 \to 2$ CC interactions with a gluon radiated off the initial or FS quarks. Here also, we find that the dominant contribution originates from the process $e u \to d g \nu_e$, which consist more than 95\% of the total $2 \to 3$ CC cross-section. 

The NC $2 \to 3$ reaction in Eq.\ref{NCeq2j} proceeds via the electron-gluon fusion processes
$e g \to q \bar q e$ and electron-quark scattering 
$e q \to q g e $ and $e  \bar q \to \bar q g e $, where $q=u,d,c,s,b$ and  $e q \to q g e $, $e  \bar q \to \bar q g e $ correspond to the NC $2 \to 2$ processes considered earlier with an additional gluon radiation off the initial and FS quarks. 
Notice that direct production of a single $b$-jet in the $2 \to 3$ NC case is possible in the SM
via $e b \to b ge$ and $e \bar b \to \bar b g e$, 
though giving a negligibly small 
contribution to the di-jet cross-section of about 0.1\% (as in the NC $2 \to 2$ case discussed earlier). 
Thus the dominant SM background contribution to $e + p/A \to j_b + j_\ell  + e$ arises from light-jet mis-identified as a $b$-jets in $e + p/A \to 2 \cdot  j  + e$, where we find that the parton-level $e u \to u g e$ is responsible for about $80\%$ of the corresponding total SM cross-section. However, as in the $2 \to 2$ case, the $2 \to 3$ NC configuration exhibits noticeably lower sensitivity to the new flavor physics considered. 

Therefore, for the di-jet case as well we concentrate below on the CC process of Eq.\ref{CCeq2j}, where here the effective SM and NP single $b$-jet cross-sections 
are (see Eq.\ref{sigmabar}): 
\begin{eqnarray}
    \bar\sigma_{1j_b}^{CC}(SM) &=& 2 t_j (1-t_j) \cdot \sigma_{SM}^{CC} ~~~~~~~~~~~~~~~~~~;~~ \sigma_{SM}^{CC} = \sigma(e + p/A \to 2 \cdot j_l + \missET) ~, \label{sigeffSM2j}\\ 
    \bar\sigma_{1j_b}^{CC}(NP) &=& \left[ \epsilon_b (1-t_j) + t_j (1-\epsilon_b) \right] \cdot \sigma_{NP}^{CC} ~~; ~~ \sigma_{NP}^{CC} = \sigma(e + p/A \to j_b + j_\ell + \missET) ~,\label{sigeffNP2j}
\end{eqnarray}
and we find that $\sigma_{SM}^{CC} \sim 109.5$fb (dominated by $e  u \to d  g \nu_e$, as noted above) and e.g.,  $\sigma_{NP}^{CC} \sim 1.6$fb for the tensor  operator ${\cal O}_{\ell e qu}^{(3)}(1131)$,   
at the EIC with beam energies of $E_e=18$ GeV and $E_p=275$ GeV.
The total effective cross-section for the $2 \to 3$ CC single $b$-jet production process $e + p/A \to j_\ell + j_b + \missET$  is given by the sum of the SM and NP contributions, i.e., $\bar\sigma_{1j_b}^{CC}(\epsilon_b,t_j,\Lambda_{\tt eff})$ of Eq.\ref{sigeff}, where the term $\propto \epsilon_b (1-t_j)$ dominates the NP part $\bar\sigma_{1j_b}^{CC}(NP)$ when $t_j \ll \epsilon_b$.

in Fig.\ref{fig:4} we depict the 95\% CL bound 
on $\Lambda_{\tt eff}$ that can be expected from a search of single $b$-jet events in di-jet signals at the EIC $e + p/A \to 2 \cdot j + \missET$, as a function of $P_e$ for the tensor operator, $\op_{\ell e q u}^{(3)}(1131) $. Here also we present results for the "loose", "medium" and "tight" $b$-tagging scenarios defined above. Evidently, the di-jet CC signal analyzed in this section is less sensitive to the scale of the FC operator $\mathcal{O}_{\ell equ}^{(3)}(1131)$ than the single-jet process discussed previously. This lower sensitivity also persists for the other FC operators examined throughout this study.

\begin{figure}[htb]
  \centering
\includegraphics[width=0.6\textwidth]{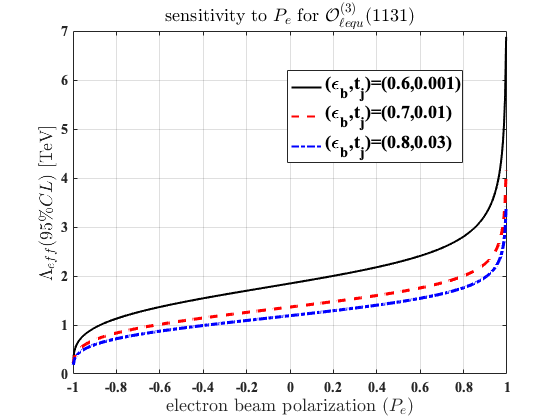}
\caption{Expected 95\% CL bounds on $\Lambda_{\tt eff}$ [TeV] from 
the CC di-jet production process, $e + p/A \to 2 \cdot j + \missET$, 
as a function of the 
electron beam polarization, $P_e$, for 
the "loose", "medium" and "tight" $b$-tagging scenarios, for the tensor operator $\op_{\ell e q u}^{(3)}(1131) $.}
  \label{fig:4}
\end{figure}

\section{Summary}

We suggest that single and di-jet production at the EIC via the CC interactions  $e +p/A \to j + \missET$ and $e +p/A \to 2 \cdot j + \missET$ ($j=j_\ell$ and $j=j_b$ for a light and $b$-quark jet, respectively) will be a valuable environment for testing new BSM TeV-scale sources of flavor violation involving the 3rd generation quarks. 
In particular, these reactions are  
even under $b$-Parity in the SM, defined by $b_P = (-1)^{n}$ \cite{our_bP_paper}, where $n$ is the number of b-quarks jets in the final state, 
since the $b_P$-odd  signals of single $b$-jet events (i.e., $b_P=-1$ for $n=1$), $e +p/A \to j_b + \missET$ and $e +p/A \to  j_b + j_l + \missET$, are necessarily proportional to the off-diagonal CKM factors $|V_{ub}|^2$ and/or $|V_{cb}|^2$ in the SM and are, therefore, un-observably small (more precisely, in the limit  $V_{j3},V_{3j} \to 0,\,j \neq 3$, the SM possesses an additional global $U(1)_b$ symmetry, "bottomness", that holds to any order in perturbation theory).    
Thus, the only significant source of SM background to $b_P$-violation (for our purpose in the CC reactions $e +p/A \to j + \missET$ and $e +p/A \to 2 \cdot j + \missET$) arises from mis-identifying a light-jet for a $b$-jet, i.e., from non-ideal purity of the $b$-jet sample.   

Conversely, we show that new types of TeV-scale flavor physics involving the 3rd generation quark fields can mediate the $b_P$-odd single $b$-jet $j_b + \missET$ and $j_b + j_\ell + \missET$ events at the EIC via
CC interactions, where we use the SMEFT framework and examine the sensitivity of the EIC to dim.6 flavor-changing $\ell \ell q q^\prime$ 4-Fermi contact terms as well operators that modify the $Wqq^\prime$ vertex; both types involving $b \to u$ transitions. In particular, since the SM is essentially $b_P$-even, we show that $b$-jet counting offers a useful  discriminator for new flavor physics effects, where the $b$-tagging efficiency and, more importantly, the purity of the $b$-jet sample, become critical factors for this type of analysis. 

Capitalizing on the different behaviors of the SM and the new flavor physics under $b$-Parity, we find that by counting single $b$-jet events in CC reactions at the EIC, one will be able to probe new flavor-violating physics with a scale
in the range $\Lambda \sim 1 - 5$, depending on the underlying heavy dynamics; 
notably, reaching a sensitivity to NP scales which are $10 - 30$ times larger than the EIC CM energy considered ($E_{CM} \sim 140$ GeV). This represents a substantial improvement over what could have been achieved at the HERA 
$ep$ collider and it will require a high purity of the $b$-jet sample and, for some types of NP, also a highly polarized electron beam. 


\bibliographystyle{hunsrt.bst}
\bibliography{mybib2}

\begin{thebibliography}{10}

\bibitem{our_bP_paper}
S.~Bar-Shalom and J.~Wudka.
\newblock {Counting inclusive b jets as an efficient probe of new flavor physics}.
\newblock {\em Phys. Rev. Lett.}, 86:3722--3725, 2001, hep-ph/9904365.

\bibitem{EIC1}
A.~Accardi et~al.
\newblock {Electron Ion Collider: The Next QCD Frontier}: {Understanding the glue that binds us all}.
\newblock {\em Eur. Phys. J. A}, 52(9):268, 2016, 1212.1701.

\bibitem{EIC2}
R.~Abdul~Khalek et~al.
\newblock {Science Requirements and Detector Concepts for the Electron-Ion Collider}: {EIC Yellow Report}.
\newblock {\em Nucl. Phys. A}, 1026:122447, 2022, 2103.05419.

\bibitem{1302.6263}
K.~S. Kumar, Sonny Mantry, W.~J. Marciano, and P.~A. Souder.
\newblock {Low Energy Measurements of the Weak Mixing Angle}.
\newblock {\em Ann. Rev. Nucl. Part. Sci.}, 63:237--267, 2013, 1302.6263.

\bibitem{Kumar:2016mfi}
K.~S. Kumar, A.~Deshpande, J.~Huang, S.~Riordan, and Y.~X. Zhao.
\newblock {Electroweak and BSM Physics at the EIC}.
\newblock {\em EPJ Web Conf.}, 112:03004, 2016.

\bibitem{1612.06927}
Y.~X. Zhao, A.~Deshpande, J.~Huang, K.~S. Kumar, and S.~Riordan.
\newblock {Neutral-Current Weak Interactions at an EIC}.
\newblock {\em Eur. Phys. J. A}, 53(3):55, 2017, 1612.06927.

\bibitem{2107.02134}
Bin Yan, Zhite Yu, and C.~P. Yuan.
\newblock {The anomalous Zbb{\textasciimacron} couplings at the HERA and EIC}.
\newblock {\em Phys. Lett. B}, 822:136697, 2021, 2107.02134.

\bibitem{2112.07747}
Hai~Tao Li, Bin Yan, and C.~P. Yuan.
\newblock {Jet charge: A new tool to probe the anomalous Zbb{\textasciimacron} couplings at the EIC}.
\newblock {\em Phys. Lett. B}, 833:137300, 2022, 2112.07747.

\bibitem{2204.07557}
Radja Boughezal, Alexander Emmert, Tyler Kutz, Sonny Mantry, Michael Nycz, Frank Petriello, Ka{\u{g}}an {\c{S}}im{\c{s}}ek, Daniel Wiegand, and Xiaochao Zheng.
\newblock {Neutral-current electroweak physics and SMEFT studies at the EIC}.
\newblock {\em Phys. Rev. D}, 106(1):016006, 2022, 2204.07557.

\bibitem{1006.5063}
Matthew Gonderinger and Michael~J. Ramsey-Musolf.
\newblock {Electron-to-Tau Lepton Flavor Violation at the Electron-Ion Collider}.
\newblock {\em JHEP}, 11:045, 2010, 1006.5063.
\newblock [Erratum: JHEP 05, 047 (2012)].

\bibitem{1401.6199}
Jens Erler, Charles~J. Horowitz, Sonny Mantry, and Paul~A. Souder.
\newblock {Weak Polarized Electron Scattering}.
\newblock {\em Ann. Rev. Nucl. Part. Sci.}, 64:269--298, 2014, 1401.6199.

\bibitem{2004.00748}
Radja Boughezal, Frank Petriello, and Daniel Wiegand.
\newblock {Removing flat directions in standard model EFT fits: How polarized electron-ion collider data can complement the LHC}.
\newblock {\em Phys. Rev. D}, 101(11):116002, 2020, 2004.00748.

\bibitem{2102.06176}
Vincenzo Cirigliano, Kaori Fuyuto, Christopher Lee, Emanuele Mereghetti, and Bin Yan.
\newblock {Charged Lepton Flavor Violation at the EIC}.
\newblock {\em JHEP}, 03:256, 2021, 2102.06176.

\bibitem{2112.02477}
Yandong Liu and Bin Yan.
\newblock {Searching for the axion-like particle at the EIC*}.
\newblock {\em Chin. Phys. C}, 47(4):043113, 2023, 2112.02477.

\bibitem{2112.04513}
Hooman Davoudiasl, Roman Marcarelli, and Ethan~T. Neil.
\newblock {Lepton-flavor-violating ALPs at the Electron-Ion Collider: a golden opportunity}.
\newblock {\em JHEP}, 02:071, 2023, 2112.04513.

\bibitem{2203.01510}
Bin Yan.
\newblock {Probing the dark photon via polarized DIS scattering at the HERA and EIC}.
\newblock {\em Phys. Lett. B}, 833:137384, 2022, 2203.01510.

\bibitem{2203.13199_Snowmass}
R.~Abdul~Khalek et~al.
\newblock {Snowmass 2021 White Paper: Electron Ion Collider for High Energy Physics}.
\newblock 3 2022, 2203.13199.

\bibitem{2207.10261}
J.~L. Zhang et~al.
\newblock {Search for e{\textrightarrow}{\ensuremath{\tau}} charged lepton flavor violation at the EIC with the ECCE detector}.
\newblock {\em Nucl. Instrum. Meth. A}, 1053:168276, 2023, 2207.10261.

\bibitem{2210.09287}
Brian Batell, Tathagata Ghosh, Tao Han, and Keping Xie.
\newblock {Heavy neutral leptons at the Electron-Ion Collider}.
\newblock {\em JHEP}, 03:020, 2023, 2210.09287.

\bibitem{2301.02304}
Radja Boughezal, Daniel de~Florian, Frank Petriello, and Werner Vogelsang.
\newblock {Transverse spin asymmetries at the EIC as a probe of anomalous electric and magnetic dipole moments}.
\newblock {\em Phys. Rev. D}, 107(7):075028, 2023, 2301.02304.

\bibitem{2307.00102}
Hooman Davoudiasl, Roman Marcarelli, and Ethan~T. Neil.
\newblock {Displaced signals of hidden vectors at the Electron-Ion Collider}.
\newblock {\em Phys. Rev. D}, 108(7):075017, 2023, 2307.00102.

\bibitem{2310.08827}
Reuven Balkin, Or~Hen, Wenliang Li, Hongkai Liu, Teng Ma, Yotam Soreq, and Mike Williams.
\newblock {Probing axion-like particles at the Electron-Ion Collider}.
\newblock {\em JHEP}, 02:123, 2024, 2310.08827.

\bibitem{2401.08419}
Hao-Lin Wang, Xin-Kai Wen, Hongxi Xing, and Bin Yan.
\newblock {Probing the four-fermion operators via the transverse double spin asymmetry at the Electron-Ion Collider}.
\newblock {\em Phys. Rev. D}, 109(9):095025, 2024, 2401.08419.

\bibitem{2402.17821}
Hooman Davoudiasl, Roman Marcarelli, and Ethan~T. Neil.
\newblock {Flavor-violating ALPs, electron g-2, and the Electron-Ion Collider}.
\newblock {\em Phys. Rev. D}, 109(11):115013, 2024, 2402.17821.

\bibitem{2411.13497}
Filippo Delzanno, Kaori Fuyuto, Sergi Gonz{\`a}lez-Sol{\'\i}s, and Emanuele Mereghetti.
\newblock {Global analysis of {\ensuremath{\mu}} {\textrightarrow} e interactions in the SMEFT}.
\newblock {\em JHEP}, 07:283, 2025, 2411.13497.

\bibitem{2503.02605}
Yongjie Deng, Xu-Hui Jiang, Tianbo Liu, and Bin Yan.
\newblock {Testing lepton flavor universality at the Electron-Ion Collider}.
\newblock {\em JHEP}, 06:157, 2025, 2503.02605.

\bibitem{2505.08871}
Hooman Davoudiasl and Hongkai Liu.
\newblock {Electron-ion collider as a discovery tool for invisible dark bosons}.
\newblock {\em Phys. Rev. D}, 112(7):075001, 2025, 2505.08871.

\bibitem{2507.02039}
Luigi Bellafronte, Sally Dawson, Pier~Paolo Giardino, and Hongkai Liu.
\newblock {Probing Top-Quark{\textendash}Electron Interactions at Future Colliders}.
\newblock {\em Phys. Rev. Lett.}, 135(25):251801, 2025, 2507.02039.

\bibitem{2507.21477}
Xu-Hui Jiang, Yiming Liu, and Bin Yan.
\newblock {Probing top-quark electroweak couplings indirectly at the Electron-Ion Collider}.
\newblock 7 2025, 2507.21477.

\bibitem{2512.15865}
Hooman Davoudiasl, Hongkai Liu, Sonny Mantry, and Ethan~T. Neil.
\newblock {Weak Charge Form Factor Determination at the Electron-Ion Collider}.
\newblock 12 2025, 2512.15865.

\bibitem{2601.00068}
Reuven Balkin, Ta'el Coren, Alexander Jentsch, Hongkai Liu, Maksym Ovchynnikov, Yotam Soreq, and Sokratis Trifinopoulos.
\newblock {Braking protons at the EIC: from invisible meson decay to new physics searches}.
\newblock 12 2025, 2601.00068.

\bibitem{HERA1}
M.~Klein and R.~Yoshida.
\newblock {Collider Physics at HERA}.
\newblock {\em Prog. Part. Nucl. Phys.}, 61:343--393, 2008, 0805.3334.

\bibitem{HERA2}
David~M. South and Monica Turcato.
\newblock {Review of Searches for Rare Processes and Physics Beyond the Standard Model at HERA}.
\newblock {\em Eur. Phys. J. C}, 76(6):336, 2016, 1605.03459.

\bibitem{EFT1}
W.~Buchmuller and D.~Wyler.
\newblock {Effective Lagrangian Analysis of New Interactions and Flavor Conservation}.
\newblock {\em Nucl. Phys.}, B268:621--653, 1986.

\bibitem{EFT2}
C.~Arzt, M.~B. Einhorn, and J.~Wudka.
\newblock {Patterns of deviation from the standard model}.
\newblock {\em Nucl. Phys.}, B433:41--66, 1995, hep-ph/9405214.

\bibitem{EFT3}
Martin~B. Einhorn and Jose Wudka.
\newblock {The Bases of Effective Field Theories}.
\newblock {\em Nucl. Phys.}, B876:556--574, 2013, 1307.0478.

\bibitem{EFT4}
B.~Grzadkowski, M.~Iskrzynski, M.~Misiak, and J.~Rosiek.
\newblock {Dimension-Six Terms in the Standard Model Lagrangian}.
\newblock {\em JHEP}, 10:085, 2010, 1008.4884.

\bibitem{EFT5}
Ilaria Brivio and Michael Trott.
\newblock {The Standard Model as an Effective Field Theory}.
\newblock {\em Phys. Rept.}, 793:1--98, 2019, 1706.08945.

\bibitem{jose_PTG_LG}
Martin~B. Einhorn and Jose Wudka.
\newblock {The Bases of Effective Field Theories}.
\newblock {\em Nucl. Phys. B}, 876:556--574, 2013, 1307.0478.

\bibitem{Greljo:2022jac}
Admir Greljo, Jakub Salko, Aleks Smolkovi{\v{c}}, and Peter Stangl.
\newblock {Rare b decays meet high-mass Drell-Yan}.
\newblock {\em JHEP}, 05:087, 2023, 2212.10497.

\bibitem{Allwicher:2022gkm}
Lukas Allwicher, Darius~A. Faroughy, Florentin Jaffredo, Olcyr Sumensari, and Felix Wilsch.
\newblock {Drell-Yan tails beyond the Standard Model}.
\newblock {\em JHEP}, 03:064, 2023, 2207.10714.

\bibitem{Greljo:2023bab}
Admir Greljo, Jakub Salko, Aleks Smolkovi{\v{c}}, and Peter Stangl.
\newblock {SMEFT restrictions on exclusive b {\textrightarrow} u{\ensuremath{\ell}}{\ensuremath{\nu}} decays}.
\newblock {\em JHEP}, 11:023, 2023, 2306.09401.

\bibitem{Grunwald:2023nli}
Cornelius Grunwald, Gudrun Hiller, Kevin Kr{\"o}ninger, and Lara Nollen.
\newblock {More synergies from beauty, top, Z and Drell-Yan measurements in SMEFT}.
\newblock {\em JHEP}, 11:110, 2023, 2304.12837.

\bibitem{Hiller:2025hpf}
Gudrun Hiller, Lara Nollen, and Daniel Wendler.
\newblock {Total Drell{\textendash}Yan in the flavorful SMEFT}.
\newblock {\em Eur. Phys. J. C}, 85(6):657, 2025, 2502.12250.

\bibitem{Bar-Shalom:1999dtk}
S.~Bar-Shalom and J.~Wudka.
\newblock {Flavor changing single top quark production channels at e+ e- colliders in the effective Lagrangian description}.
\newblock {\em Phys. Rev. D}, 60:094016, 1999, hep-ph/9905407.

\bibitem{L3}
P.~Achard et~al.
\newblock {Search for single top production at LEP}.
\newblock {\em Phys. Lett. B}, 549:290--300, 2002, hep-ex/0210041.

\bibitem{DELPHI}
J.~Abdallah et~al.
\newblock {Search for single top quark production via contact interactions at LEP2}.
\newblock {\em Eur. Phys. J. C}, 71:1555, 2011, 1102.4455.

\bibitem{Maltoni-global}
Gauthier Durieux, Fabio Maltoni, and Cen Zhang.
\newblock {Global approach to top-quark flavor-changing interactions}.
\newblock {\em Phys. Rev. D}, 91(7):074017, 2015, 1412.7166.

\bibitem{topdecay1}
Radja Boughezal, Chien-Yi Chen, Frank Petriello, and Daniel Wiegand.
\newblock {Top quark decay at next-to-leading order in the Standard Model Effective Field Theory}.
\newblock {\em Phys.\ Rev.\ D}, 100(5):056023, 2019, 1907.00997.

\bibitem{1008.3562}
J.A. Aguilar-Saavedra.
\newblock {Effective four-fermion operators in top physics: A Roadmap}.
\newblock {\em Nucl. Phys. B}, 843:638--672, 2011, 1008.3562.
\newblock [Erratum: Nucl.Phys.B 851, 443--444 (2011)].

\bibitem{topdecay3}
Sacha Davidson, Michelangelo~L. Mangano, Stephane Perries, and Viola Sordini.
\newblock {Lepton Flavour Violating top decays at the LHC}.
\newblock {\em Eur. Phys. J. C}, 75(9):450, 2015, 1507.07163.

\bibitem{topdecay2}
Mikael Chala, Jose Santiago, and Michael Spannowsky.
\newblock {Constraining four-fermion operators using rare top decays}.
\newblock {\em JHEP}, 04:014, 2019, 1809.09624.

\bibitem{madgraph5}
Johan Alwall, Michel Herquet, Fabio Maltoni, Olivier Mattelaer, and Tim Stelzer.
\newblock {MadGraph 5 : Going Beyond}.
\newblock {\em JHEP}, 06:128, 2011, 1106.0522.

\bibitem{SMEFTsim1}
Ilaria Brivio, Yun Jiang, and Michael Trott.
\newblock {The SMEFTsim package, theory and tools}.
\newblock {\em JHEP}, 12:070, 2017, 1709.06492.

\bibitem{SMEFTsim2}
Ilaria Brivio.
\newblock {SMEFTsim 3.0 \textemdash{} a practical guide}.
\newblock {\em JHEP}, 04:073, 2021, 2012.11343.

\bibitem{1912.10053}
Tie-Jiun Hou et~al.
\newblock {New CTEQ global analysis of quantum chromodynamics with high-precision data from the LHC}.
\newblock {\em Phys. Rev. D}, 103(1):014013, 2021, 1912.10053.

\end{thebibliography}

\end{document}